\documentclass{elsart1p}

\usepackage{graphics}
\usepackage{graphicx}
\usepackage{bm}
\usepackage{epsfig}
\usepackage{amsmath,amssymb}

\usepackage[normalem]{ulem}  
\usepackage{color} 

\renewcommand\sout{\bgroup \color{red} \ULdepth=-.5ex \ULset}

\begin{document}

\begin{flushright}
KEK-TH-1621 \\ J-PARC-TH-0025
\end{flushright}

\begin{frontmatter}

\title{Spin degeneracy in multi-hadron systems \\ with a heavy quark}

\author[a]{Shigehiro Yasui\corauthref{cor}},
\ead{yasuis@post.kek.jp}
\author[b]{Kazutaka Sudoh},
\author[c]{Yasuhiro Yamaguchi},
\author[c]{Shunsuke Ohkoda}, 
\author[c]{Atsushi Hosaka},
\author[d]{Tetsuo Hyodo\thanksref{label1}}
\thanks[label1]{The present address is Yukawa Institute for Theoretical Physics, Kyoto University, Kyoto, 606-8317, Japan.}

\address[a]{KEK Theory Center, Institute of Particle and Nuclear
Studies, High Energy Accelerator Research Organization, 1-1, Oho,
Ibaraki, 305-0801, Japan}

\address[b]{Nishogakusha University, 6-16, Sanbancho, Chiyoda,
Tokyo, 102-8336, Japan}

\address[c]{Research Center for Nuclear Physics (RCNP), 
Osaka University, Ibaraki, Osaka, 567-0047, Japan}

\address[d]{Department of Physics, Tokyo Institute of Technology,
Tokyo 152-8551, Japan}

\corauth[cor]{Corresponding author at: KEK Theory Center, Institute of Particle and Nuclear
Studies, High Energy Accelerator Research Organization, 1-1, Oho,
Ibaraki, 305-0801, Japan}



\begin{abstract}
We study multi-hadron systems with a single heavy quark (charm or bottom) in the limit of heavy quark mass.
The spin degeneracy of the states with quantum numbers $(j+1/2)^{P}$ and $(j-1/2)^{P}$ for $j \neq 0$, known in a normal hadron, can be generalized to multi-hadron systems. 
The spin degeneracy is the universal phenomena for any multi-hadron systems with a single heavy quark, irrespective of their internal structures, including  compact multi-quarks, hadronic molecules and exotic nuclei.
We demonstrate 
 the spin degeneracy
 in the hadronic systems formed by a heavy hadron effective theory; $P^{(\ast)}N$ states with a $P^{(\ast)}=\bar{D}^{(\ast)}$, $B^{(\ast)}$ meson and a nucleon $N$, and a $P^{(\ast)}$ meson in nuclear matter.
\end{abstract}
\end{frontmatter}

Recent experimental developments in hadron spectroscopy have unveiled the existence of various exotic hadrons which are considered to have  extraordinary structures.
To analyze their properties is intimately related to the fundamental problems in QCD, such as color confinement and dynamical chiral symmetry breaking.
Especially for charm and bottom flavors, there have been many experimental 
evidences for
 the existence of the exotic hadrons, such as $X$, $Y$ and $Z$ for charm sector and $Y_b$ and $Z_b$ for bottom sector~\cite{Brambilla:2010cs}.
Theoretically, not only exotic hadrons, but also exotic nuclei with charm and bottom are discussed.
Those states can be explored experimentally at facilities, such as J-PARC, GSI-FAIR, RHIC and LHC~\cite{Cho:2010db}.
Heavy exotic hadrons and nuclei will bring us a new insight from the quark dynamics to the nuclear dynamics, which cannot be accessed by light flavor hadrons.
Although many model calculations have been extensively performed in the literature, it will be eagerly required to have the rigorous knowledge directly based on QCD.

A unique feature of charm and bottom quarks is that their masses are heavier than the energy scale of light quark QCD.
In the heavy mass limit, it leads to the spin symmetry~\cite{Isgur:1989vq,Isgur:1991wq,Rosner:1985dx,Eichten:1993ub,Manohar:2000dt}.
It has been known that, in a hadron with a single heavy quark, the spin of the heavy quark spin is decoupled from the total angular momentum of the light quarks and gluons, the ``brown muck" which is everything other than the heavy quark in the hadron~\cite{Flynn:1992fm}.
As a consequence, there appears a pair of degenerate states with total angular momentum and parity, $(j-1/2)^{P}$ and $(j+1/2)^{P}$, for $j\neq 0$, while a single state for $j=0$, where $j$ is the total angular momentum of the brown muck.
The spin degeneracy has been addressed in the context of normal hadrons, such as mesons and baryons including excited states~\cite{Isgur:1991wq,Rosner:1985dx,Eichten:1993ub,Manohar:2000dt,Flynn:1992fm,Bardeen:2003kt,Matsuki:2007zza,Roberts:2007ni}.

The purpose of the present Letter is to apply the idea of the spin degeneracy to multi-hadron systems such as exotic hadrons (irrespective of multi-quarks, hadronic molecules and exotic nuclei) containing a single heavy quark.
We investigate the spin degeneracy in the hadronic effective theory of QCD, where the fundamental degrees of freedom are given by hadrons, and show that hadronic molecules with baryon number one and heavy mesons in nuclear matter exhibit the spin degeneracy.
Throughout the discussion, we assume non-negative baryon numbers for the multi-hadron systems.
The cases of negative baryon numbers will be immediately obtained.

First of all, let us consider the spin degeneracy in multi-hadron systems in views of QCD.
This is a general discussion so that the conclusion should hold in any multi-hadron systems, as far as the heavy quark limit is adopted.
Denoting the four velocity of the heavy quark as $v^{\mu}$ with $v^2=1$,
we introduce
the effective field
$Q_{v}(x) = e^{i m_{\rm Q} v\cdot x} \frac{1+v\hspace{-0.5em}/}{2} Q(x)$
for projecting out the positive energy component of the heavy quark field $Q(x)$.
The effective Lagrangian in the $1/m_{\rm Q}$ expansion is given by
\begin{eqnarray}
{\cal L}_{\rm{HQET}} = \bar{Q}_{v} v \cdot i D Q_{v} + {\cal O}(1/m_{\rm Q}),
\end{eqnarray} 
with the covariant derivative $D_{\mu}$.
This is the Lagrangian with a leading term
 in the heavy quark effective theory (HQET)~\cite{Isgur:1989vq,Isgur:1991wq,Rosner:1985dx,Eichten:1993ub,Manohar:2000dt}.
A hadron with a single heavy quark is composed of the heavy quark with spin $\vec{S}$ and the brown muck with total angular momentum $\vec{j}$.
The total angular momentum of the hadron is $\vec{J} = \vec{S} + \vec{j}$.
In the limit of heavy quark mass ($m_{\mathrm Q}  \rightarrow  \infty$), $\vec{S}$ is a conserved quantity, because the spin flip terms are absent in the leading order in the $1/m_{\rm Q}$ expansion of ${\cal L}_{\rm{HQET}}$.
This is called the heavy quark spin (HQS) symmetry.
Clearly $\vec{J}$ is conserved.
Therefore, we find that $\vec{j}$ is conserved, even though the brown muck is a highly non-perturbative object.
Thus, we confirm that the spin degeneracy is realized as addressed previously.

Interestingly, the notion of the spin degeneracy is generally applied, 
not only to normal hadrons in ground states and higher exited states, but also to multi-hadron systems and even to exotic nuclear systems, as far as the states contain a single heavy (anti)quark.
Here we consider exotic hadrons with a single heavy antiquark,
whose quark contents are minimally given by $\bar{Q}q^n$ ($n=3B+1$ with baryon number $B\ge0$) with a heavy antiquark $\bar{Q}$ and many light quarks $q$.
We should note that 
the state is in fact a superposition of $n$ light quarks plus any number of $q\bar{q}$ pairs and gluons $g$ for a given quantum number;
\begin{eqnarray}
&& \bar{Q}\underbrace{q \cdots q} + \bar{Q}\underbrace{q \cdots q}q\bar{q} + \bar{Q}\underbrace{q \cdots q}q\bar{q}g + \cdots. \label{eq:quark_configuration} \\
&& \hspace{1.9em}n \hspace{3.7em}n \hspace{5.3em}n \nonumber
\end{eqnarray}
Then, we ask the question whether the exotic hadrons have the spin degeneracy as normal hadrons.
In those systems, the heavy quark spin for $\bar{Q}$ is a conserved quantity in the heavy mass limit, and the total angular momentum is also conserved.
Therefore, the total angular momentum of the ensemble of the light components, $q \cdots q + q \cdots qq\bar{q} + q \cdots qq\bar{q}g + \cdots$, in Eq.~(\ref{eq:quark_configuration}) is also conserved.
Consequently, we obtain the result that there is a pair of degenerate states with $(j-1/2)^{P}$ and $(j+1/2)^{P}$ for $j\neq 0$, and a single state for $j=0$, where $j^{\cal P}$ is the total angular momentum and parity of the light components (${\cal P}=-P$).
Hereafter we call those states ``HQS doublets" for $j\neq 0$ and ``HQS singlets" for $j=0$, respectively.

In the present discussion, we call the light components in Eq.~(\ref{eq:quark_configuration}) ``light spin-complex" (or ``spin-complex" in short).
The reason for introducing the new term is explained.
When the state is a compact multi-quark, the spin-complex is an ensemble of light quarks and gluons.
This is 
the ordinary situation for
 the brown muck in a normal hadron.
When the state is a {\it spatially extended hadronic molecule}, however, the state can be composed of light hadrons and the $\bar{Q}q$ meson.
Then, the spin-complex is a composite system with light hadrons and 
light components in the $\bar{Q}q$ meson.
Here we use ``spin" to emphasize the importance of spin degrees of freedom. 
As a typical configuration, 
 it can be $q\bar{q}$ mesons and a light quark $q$ for $B=0$, or $qqq$ baryons and a light quark $q$ for $B\ge 1$.
The latter can be applied to the exotic nuclei containing the $\bar{Q}q$ meson, as discussed later.
We discriminate the spin-complex from the conventional brown muck, in the sense that the spin-complex describes the composite systems by quarks, gluons and hadrons.

How the spin-complex is formed 
 is a highly non-perturbative problem in QCD.
However, the spin-complex is a useful object to classify exotic hadrons,
when it carries a good quantum number $j^{\cal P}$.
Therefore, the spin (non-)degeneracy for $j\neq 0$ ($j=0$)
 is a universal phenomena regardless whether the state is a compact multi-quark state, an extended hadronic molecule state or even a mixture of them.

From the above discussions, the spin degeneracy can occur in any multi-hadron systems with a single heavy quark, such as hadronic molecules and nuclei, in the heavy quark limit.
However, it is not known a priori how the spin degeneracy occurs 
in the {\it heavy hadron effective theory}.
Specifically, it is not a trivial problem whether the ground state is a HQS doublet or a HQS singlet.
This is the main subject which will be investigated as follows.

To start with, let us discuss hadronic molecules with genuinely exotic quark content $\bar{Q}qqqq$.
We assume that they are compound by a heavy meson $P^{(\ast)} \sim \bar{Q}q$ and a nucleon $N$.
We denote the heavy meson with $J^{P}=0^{-}$ ($1^{-}$) as $P$ ($P^{\ast}$), and $P^{(\ast)}$ stands for one of the HQS doublet $(P,P^{\ast})$.
We study whether the $P^{(\ast)}N$ systems exhibit the spin degeneracy in terms of the heavy meson effective theory.

To demonstrate concretely, we consider the one pion exchange potential between $P^{(\ast)}$ and $N$ in Refs.~\cite{Yasui:2009bz,Yamaguchi:2011xb}.
The heavy quark symmetry is realized in the degenerate masses of $P$ and $P^{\ast}$ and the common couplings of $\pi PP^{\ast}$ and $\pi P^{\ast} P^{\ast}$ vertices.
We will investigate the states with $1/2^{-}$ and $3/2^{-}$ in detail, and 
later will extend the discussion to general cases of $(j-1/2)^{P}$ and $(j+1/2)^{P}$.

The wave function in $P^{(\ast)}N$
 has multi-channels;
\begin{eqnarray}
\left\{ | PN(^2{\mathrm S}_{1/2}) \rangle, | P^\ast \! N(^2{\mathrm S}_{1/2}) \rangle, | P^\ast \! N(^4{\mathrm D}_{1/2}) \rangle \right\},
\end{eqnarray}
 for $1/2^{-}$ and
\begin{eqnarray}
\left\{ | PN(^2{\mathrm D}_{3/2}) \rangle, \! | P^\ast \! N(^4{\mathrm S}_{3/2}) \rangle, \! | P^\ast \! N(^4{\mathrm D}_{3/2}) \rangle, \! | P^\ast \! N(^2{\mathrm D}_{3/2}) \rangle \right\},
\end{eqnarray}
 for $3/2^-$.
Then, the Hamiltonians are given by
\begin{eqnarray}
H_{1/2^-} \!&=&\!
\left(
\begin{array}{ccc}
 K_0 & \sqrt{3} \, {C} & -\sqrt{6} \, {T}  \\
\sqrt{3} \, {C} & K_0-2 \, {C} & -\sqrt{2} \, {T} \\
-\sqrt{6} \, {T} & -\sqrt{2} \, {T} & K_2 + ({C} - 2\, {T})
\end{array}
\right), \\
H_{3/2^-} \!&=&\!
\left(
\begin{array}{cccc}
 K_2 & \sqrt{3}\, {T} & -\sqrt{3} \, {T} & \sqrt{3}\,{C} \\
\sqrt{3}\,{T} &K_0 + {C} & 2\,{T} & {T} \\
-\sqrt{3}\,{T} & 2\,{T} & K_2 + {C} & -{T} \\
\sqrt{3}\,{C} & {T} & -{T} & K_2 -2\,{C}
\end{array}
\right), 
\end{eqnarray}
for $1/2-$ and $3/2^-$, respectively.
We define the kinetic term $K_{\ell}=-( \partial^2 / \partial r^2 + (2/r)\partial / \partial r - \ell(\ell+1)/r^2)/2\mu$ for angular momentum $\ell$, and the reduced mass $\mu$ coincides with the mass of the nucleon $m_{N}$ in the limit of infinite mass of $P^{(\ast)}$.
We also define ${C}=\kappa\, C(r;m_\pi)$ and ${T} = \kappa\, T(r;m_\pi)$, with $\kappa = (g_{\pi}g_{\pi NN}/\sqrt{2} m_N f_{\pi})
(\vec{\tau}_{P} \!\cdot\! \vec{\tau}_N/3)$ with a $P^{(\ast)}P^{\ast}\pi$ ($NN\pi$) coupling constant $g_{\pi}$ ($g_{\pi NN}$), the pion decay constant $f_{\pi}$, and isospin matrices $\vec{\tau}_{P}$ and $\vec{\tau}_{N}$ for $P^{(\ast)}$ and $N$, respectively, and the central potential $C(r;m_\pi)$ and the tensor potential $T(r;m_\pi)$ ($r$ the distance between $P^{(\ast)}$ and $N$, and $m_\pi$ the pion mass) whose explicit forms are given in Refs.~\cite{Yasui:2009bz,Yamaguchi:2011xb}.
The $PN$ and $P^{\ast}N$ states can be mixed, and the states with different angular momenta can also be mixed.
The former originates in the heavy quark spin symmetry,
and the latter does in the tensor force of the pion exchange potential.

Previously we mentioned the existence of the spin-complex in the multi-hadron systems.
We then ask what the spin-complex in the present $P^{(\ast)}N$ system is.
We remember that the $P^{(\ast)}N$ molecule state is composed of the meson $P^{(\ast)} \sim \bar{Q}q$ and a nucleon $N$.
In view of the heavy quark spin symmetry, this system is decomposed into the heavy antiquark $\bar{Q}$ and the remaining  light degrees of freedom, namely the spin-complex.
In our model space, the spin-complex is given by $Nq$ from the nucleon $N$ and the light quark $q$ in $P^{(\ast)}$.
Combined with the heavy antiquark $\bar{Q}$, the basis states can be written as
\begin{eqnarray}
\left\{ | [Nq]^{(0,\mathrm S)}_{0^+} \, \bar{Q} \rangle_{1/2^-}, | [Nq]^{(1, {\mathrm S})}_{1^+} \bar{Q} \rangle_{1/2^-}, | [Nq]^{(1,{\mathrm D})}_{1^+} \bar{Q} \rangle_{1/2^-} \right\},
\end{eqnarray}
and 
\begin{eqnarray}
\left\{ | [Nq]^{(1, {\mathrm S})}_{1^+} \bar{Q} \rangle_{3/2^-}, | [Nq]^{(1, {\mathrm D})}_{1^+} \bar{Q} \rangle_{3/2^-},  | [Nq]^{(0, {\mathrm D})}_{2^+}  \bar{Q} \rangle_{3/2^-},  | [Nq]^{(1,{\mathrm D})}_{2^+} \bar{Q} \rangle_{3/2^-} \right\},
\end{eqnarray}
respectively, for $1/2^{-}$ and $3/2^{-}$. 
Here $[Nq]^{(S,L)}_{j^{\cal P}}$ denotes the spin-complex composed of a nucleon and a quark (in $P^{(\ast)}$ meson) with total spin $S$, angular momentum $L$, and total angular momentum and parity $j^{\cal P}$.
The particle basis by $\{ |P^{(\ast)}N(^{2S+1}L_{J})\rangle \}$ and the spin-complex basis by $\{ | [Nq]^{(S,L)}_{j^{\cal P}} \bar{Q} \rangle_{J^{P}} \}$ are related by unitary matrix, $U$ for $1/2^-$ and $U'$ for $3/2^-$.
In the spin complex basis, the Hamiltonians $H_{1/2^-}$ and $H_{3/2^-}$ are transformed as
\begin{eqnarray}
H_{1/2^-}^{\mathrm{SC}}
&\equiv& U^{-1} H_{1/2^-} U \nonumber \\
&=&
\left(
\begin{array}{c|cc}
 K_{0} \!-\! 3\,{C} & 0 & 0 \\
 \hline
 0 & K_{0} \!+\! {C} & -2\sqrt{2} \,{T} \\
 0 & -2\sqrt{2} \,{T} & K_{2} \!+\!  ({C} \!-\! 2\,{T})
\end{array}
\right) \nonumber \\
&\equiv&
 {\mathrm{diag}}\left( H_{1/2^-}^{\mathrm{SC}(0^+)},H_{1/2^-}^{\mathrm{SC}(1^+)} \right), \\
H_{3/2^-}^{\mathrm{SC}}
&\equiv&
 U'^{-1} H_{3/2^-} U' \nonumber \\
&=&
\left(
\begin{array}{cc|cc}
 K_{0} \!+\! {C} & 2\sqrt{2}\,{T} & 0 & 0 \\
 2\sqrt{2}\,{T} & K_{2} \!+\! ({C} \!-\! 2\,{T}) & 0 & 0 \\
\hline
 0 & 0 & K_{2} \!-\! 3\,{C} & 0 \\
 0 & 0 & 0 & K_{2} \!+\! ({C} \!+\! 2\,{T})
\end{array}
\right) \nonumber \\
&\equiv&
{\mathrm{diag}}\left( H_{3/2^-}^{\mathrm{SC}(1^+)},H_{3/2^-}^{\mathrm{SC}(2^+)} \right).
\end{eqnarray}
Thus, with the spin-complex basis, we obtain the block-diagonal forms
with the notation $H_{J^{P}}^{\mathrm{SC}(j^{\cal P})}$ for the $J^{P}$ state containing the spin-complex with $j^{\cal P}$.
We comment that the off-diagonal terms in $H^{\mathrm{SC}(2^{+})}_{3/2^{-}}$ vanish because the one pion exchange potential is used. 
When other interactions are employed,
 the off-diagonal terms in $H_{J^P}^{\mathrm{SC}(j^{\cal P})}$ may exist in general.
In our model space, we consider only $Nq$ for the spin-complex.
Even when the other components such as $\Delta q$, $N \pi q$ are considered, 
the Hamiltonians are block-diagonalized with the spin-complex basis,
 as far as the heavy quark symmetry is taken.

Importantly, we note that $H_{1/2^-}^{\mathrm{SC}(1^+)}$ coincides with $H_{3/2^-}^{\mathrm{SC}(1^+)}$ except for the irrelevant sign of the off-diagonal terms.
Therefore, $H_{1/2^-}^{\mathrm{SC}(1^+)}$ and $H_{3/2^-}^{\mathrm{SC}(1^+)}$ have exactly the same eigenvalues and eigenstates.
For bound states, this is truly the spin degeneracy for $P^{(\ast)}N$ with $1/2^-$ and $3/2^-$, because the spin-complex with $1^{+}$ is commonly contained in $1/2^-$ and $3/2^-$ states.
For scattering states, the equivalence is also seen with the spin-complex basis.
Thus, we find that the $1/2^-$ and $3/2^-$ states containing a common spin-complex with $1^+$, which are described by $H_{1/2^-}^{\mathrm{SC}(1^+)}$ and $H_{3/2^-}^{\mathrm{SC}(1^+)}$ respectively, belong to the HQS doublet.
On the other hand, the $1/2^-$ state with a spin-complex with $0^{+}$ described by $H_{1/2^-}^{\mathrm{SC}(0^+)}$ is the HQS singlet. 
The block-diagonalization of Hamiltonians is possible also for higher spin states.
For example, $H_{3/2^-}^{\mathrm{SC}(2^+)}$ and $H_{5/2^-}^{\mathrm{SC}(2^+)}$ form a degenerate pair of $3/2^-$ and $5/2^-$ containing the common spin-complex with $2^+$.

The eigenstates of $H_{1/2^-}^{\mathrm{SC}(1^+)}$ and $H_{3/2^-}^{\mathrm{SC}(1^+)}$ are
 the linear combinations as
\begin{eqnarray}
 \cos\theta| [Nq]^{(1, {\mathrm S})}_{1^+} \bar{Q} \rangle_{1/2^-}+ \sin\theta| [Nq]^{(1,{\mathrm D})}_{1^+} \bar{Q} \rangle_{1/2^-},
\end{eqnarray}
and
\begin{eqnarray}
\cos\theta| [Nq]^{(1, {\mathrm S})}_{1^+} \bar{Q} \rangle_{3/2^-}+ \sin\theta| [Nq]^{(1, {\mathrm D})}_{1^+} \bar{Q} \rangle_{3/2^-},
\end{eqnarray}
respectively.
Although the mixing angle $\theta$ is determined by the interaction among the light degrees of freedom, the same mixing angle for $1/2^-$ and $3/2^-$ is required from the heavy quark symmetry.
This means that the fraction of each component in the particle basis is uniquely determined.
The fraction is related to decay and production rates in weak, electromagnetic and strong processes.

We can similarly discuss the spin degeneracy for any pair of states with $(j-1/2)^P$ and $(j+1/2)^P$ for $j\neq 0$, which contain the spin-complex with $j^{{\cal P}}$ (${\cal P}=-P$) in common.
The Hamiltonians of the $(j-1/2)^P$ and $(j+1/2)^P$ states are the same;
\begin{eqnarray}
H_{(j-1/2)^{P}}^{\mathrm{SC}(j^{{\cal P}})}= H_{(j+1/2)^{P}}^{\mathrm{SC}(j^{{\cal P}})}.
\end{eqnarray}
For higher $j\pm1/2$, there can be resonant states.
The spin degeneracy occurs, not only for the bound states,
but also for the resonant states.

Thus, we have successfully presented the existence of the spin degeneracy and the spin non-degeneracy in $P^{(\ast)}N$ from the heavy meson effective theory.
However, it is still a problem whether the ground state is either a HQS doublet or a HQS singlet.
As numerical calculations from the OPEP using the parameter values in Refs.~\cite{Yasui:2009bz,Yamaguchi:2011xb}, it turns out that the spin degenerate states with $1/2^-$ and $3/2^-$ in isosinglet are the most stable states.
Therefore, the OPEP gives a HQS doublet as the ground state of $P^{(\ast)}N$.

\begin{figure}[tbp]
\begin{center}
\includegraphics[width=6cm,angle=0]{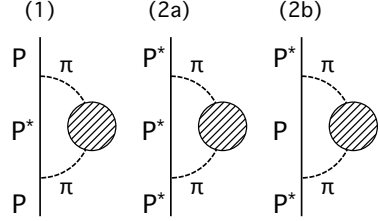}
\caption{The diagrams of the self-energies for (1) $P$ meson and (2a,b) $P^{\ast}$ meson in the nuclear matter. The solid (dashed) lines are for $P^{(\ast)}$ (pions). See the text for details.}
\label{fig:selfenergy}
\end{center}
\end{figure}

The result for the spin degeneracy found in the hadronic molecules with baryon number one ($P^{(\ast)}N$)
 will be applied to exotic nuclei with arbitrary baryon number, as far as they contain a single heavy quark.
As a limiting case, 
 we consider the uniform nuclear matter with a single $P^{(\ast)} \sim \bar{Q}q$ meson.
Note that the ground state is the nuclear matter, not vacuum.
When the $P^{(\ast)}$ meson exists in the nuclear matter (e.g. see Ref.~\cite{Yasui:2012rw} and the references therein for $\bar{D}^{(\ast)}$ ($B^{(\ast)}$) meson), the $P^{(\ast)}$ meson interacts with the neighboring nucleons, and the spin-complex may contain $NN^{-1}q+NNN^{-1}N^{-1}q+\cdots$ with the particle (nucleon) $N$, the hole $N^{-1}$ and the light quark $q$ from $P^{(\ast)}$.
Because the spin of the $\bar{Q}$ quark is irrelevant to the structure of the spin-complex, it should lead to the spin degeneracy, namely the degeneracy of the self-energies of $P$ and $P^{\ast}$ in the nuclear matter.

Let us discuss 
the self-energy of the $P^{(\ast)}$ meson in the nuclear matter, given by the diagrams in Fig.~\ref{fig:selfenergy}.
We perform the perturbative analysis at 
  ${\cal O}(g_{\pi}^2)$.
We introduce the pion self-energy $-i\Pi_{\pi}^{ab}(k)$ in the nuclear matter with the pion momentum $k$ and the final (initial) isospin $a$ ($b$), as shown by the blobs.
Indeed, it stands for the creation of many pairs of particle and hole, $NN^{-1}+NNN^{-1}N^{-1}+\cdots$, which contribute to the spin-complex.
In the heavy mass limit, the self-energies of $P$ and $P^{\ast}$
 are given, respectively, as
\begin{eqnarray}
-i\Sigma_{P} =  
&& \int \frac{{\rm d}^{4}k}{(2\pi)^{4}} \frac{k^{2} \!-\! (v \!\cdot\! k)^{2}}{2v \!\cdot\! (-k) \!+\! i \epsilon} \left( \frac{1}{k^{2} \!-\! m_{\pi}^{2} \!+\! i \epsilon} \right)^{2} \eta(k),
\end{eqnarray}
with  
\begin{eqnarray}
\eta(k)\!=\!-\!\left( \frac{2g_{\pi}}{\sqrt{2} f_{\pi}} \right)^{2} \! \tau^{a}\Pi^{ab}_{\pi}(k)\tau^{b}
\end{eqnarray}
for $P$ in Fig.~\ref{fig:selfenergy}(1), and
\begin{eqnarray}
-i\Sigma_{P^{\ast}}=-i\Sigma_{P^{\ast}}^{(P^{\ast})}-i\Sigma_{P^{\ast}}^{(P)},
\end{eqnarray}
with $\Sigma_{P^{\ast}}^{(P^{\ast})} =(2/3) \Sigma_{P}$ and $\Sigma_{P^{\ast}}^{(P)} =(1/3) \Sigma_{P}$
for $P^{\ast}$ in Fig.~\ref{fig:selfenergy}(2a,b).
Here $v^{\mu}$ is the four velocity for $P^{(\ast)}$, and $\epsilon$ is an infinitely small positive number.
The factor in the numerator in the first term of the integrand results from the spin structure of the $P^{(\ast)}$ mesons.
We note that, thanks to the heavy quark spin symmetry (i.e. the same coupling strengths for the $PP^{\ast}\pi$ and $P^{\ast}P^{\ast}\pi$ vertices and the large mass limit for $P$ and $P^{\ast}$), the contributions from (1) and (2a,b) in Fig.~\ref{fig:selfenergy} are the same except for the overall coefficients.
Their fraction $3:2:1$
 is understood intuitively from the counting of the spin degrees of freedom in the intermediate states.
Consequently, we find that the self-energy of the $P$ meson is equal to that of the $P^{\ast}$ meson;
\begin{eqnarray}
-i\Sigma_{P} = -i\Sigma_{P^{\ast}}.
\end{eqnarray}
Thus, we find that $P$ and $P^{\ast}$ in nuclear matter belong to the HQS doublet.
We confirm that the present conclusion is consistent with the result in Ref.~\cite{Yasui:2012rw}, where the pion self-energy was calculated at one-loop order, when the limit of heavy quark mass is taken.
It will be naturally expected that the spin degeneracy will hold for any higher order of $g_{\pi}$ as well as for any excited pairs of particle and hole~\cite{Yasui:2013xr}.

We straightforwardly apply our discussion to spin (non-)degeneracy in atomic nuclei with a $P^{(\ast)}$ meson contained.
Let us suppose the case that a $P^{(\ast)}$ meson is bound in $s$-wave in the parent nucleus.
When the parent nucleus has an integer spin ($0$, $1$, $2$, $\dots$), namely an even baryon number, HQS doublets can exist because the spin-complex has a half-integer spin.
There cannot exist HQS singlets.
On the other hand, when the parent nucleus has a half-integer spin ($1/2$, $3/2$, $5/2$, $\dots$), namely an odd baryon number, there can exist HQS singlets and HQS doublets, where the former contains the spin-complex with zero spin (for the parent nucleus with $1/2$) and the latter contains the spin-complex with non-zero integer spin (for the parent nucleus with $1/2$, $3/2$, $5/2$, $\dots$).
Whether a HQS singlet or a HQS doublet is realized in the ground state is determined by the interaction between a $P^{(\ast)}$ meson and a nucleon, i.e. the non-perturbative dynamics for light components in QCD.

So far we have concentrated on a heavy {\it antiquark} $\bar{Q}$,
and have presented that spin degeneracy is realized in $P^{(\ast)}N$ hadronic states and $P^{(\ast)}$ in nuclear matter.
A similar discussion will be applied to the multi-hadron systems with a heavy {\it quark} $Q$, whose quark contents are given by $Q\bar{q}$ or $Qq^{n}$ ($n=3B-1$ with baryon number $B\ge1$).
The spin degeneracy of $(j\pm1/2)^{P}$ states for $j \neq 0$ and non-degeneracy for $j=0$
 should hold for any states.
In the literature, the $D^{(\ast)}N$ ($\bar{B}^{(\ast)}N$)~\cite{Yamaguchi:2013ty} and $\Lambda_{c}N-\Sigma_{c}^{(\ast)}N$~\cite{Liu:2011xc} systems for the hadronic molecules as well as the $DNN$ systems~\cite{Bayar:2012dd} and the nuclear matter with a $D$ meson~\cite{Mizutani:2006vq} for the exotic nuclei are discussed.
Once again, in the limit of heavy quark mass, the bound states as well as resonant states should exhibit the spin degeneracy.
The most stable hadron with a heavy quark for baryon number $B = 1$ is a $\Lambda_{Q}$ baryon, such as $\Lambda_{c}$ and $\Lambda_{b}$, which is a HQS singlet.
As speculations, we may expect that the ground states of the exotic nuclei with a heavy quark are $\Lambda_{Q}$ nuclei which exhibit the HQS singlets\footnote{Strictly to say, $\Lambda_{Q}$ in nuclear medium will be dressed by nucleons and holes, and hence the quasi-particle carrying the same quantum number as $\Lambda_{Q}$ should be considered.}.
However, there is still a room that the HQS doublets could be realized, if $\Sigma_{Q}^{(\ast)}$ baryons, such as $\Sigma_{c}^{(\ast)}$ and $\Sigma_{b}^{(\ast)}$ baryons, which are HQS doublets, interact with nucleons more attractively rather than $\Lambda_{Q}$ baryons do.
It will be probable that the isospin factor from isovector nature of $\Sigma_{Q}^{(\ast)}$ baryons may enhance the attractive interaction with nucleons rather than the isospin factor from isoscalar nature of $\Lambda_{Q}$ baryons do.
It will be an interesting subject to explore the spin degeneracy and/or non-degeneracy in various multi-hadron systems with a heavy (anti)quark in future studies.

Final remarks are in order.
First, the concept of spin-complex for multi-hadron systems
 brings us a new insight for the constituents of hadrons.
In some models in the low energy QCD, the fundamental degrees of freedom are considered to be quarks in the constituent quark model, and hadrons in a hadronic molecule model.
However, the heavy quark spin symmetry tells us an importance to introduce the spin-complex as a mixture of light quarks, gluons and hadrons.
This is a new way to look especially at exotic hadrons and nuclei with a single heavy quark.
Second, the spin degeneracy
induced from the existence of the spin-complex 
 will give a guidance for experimental search of the exotic hadrons and nuclei with a single charm or bottom quark.
In the real world, however, the breaking of the heavy quark symmetry violates the strict spin degeneracy.
Nevertheless, the spin-complex is useful to both theoretical and experimental analyses.
The extrapolation from charm and bottom will enable us to reach the information in the heavy quark limit.
When one state is observed, the partner state can also be found nearby.
Furthermore the structure of the spin-complex will be studied through the decay and production rates.

In summary, we consider multi-hadron systems with a single heavy quark in the limit of heavy quark mass.
The spin degeneracy is a universal phenomena of QCD for the multi-hadrons with a single heavy (anti)quark, regardless whether the systems are the compact multi-quark states or the extended hadronic molecule states or the mixture of them.
We also show explicitly based on the heavy hadron effective theory that the spin degeneracy
 is realized for the $P^{(\ast)}N$ hadronic molecule systems and the $P^{(\ast)}$ meson in the nuclear matter.
We propose the existence of the spin-complex as a new object in the multi-hadron systems.
The notion of the spin-complex may be applied further in the deconfinement phases, such as the quark-gluon plasma, the color superconductivity and the quarkyonic matter~\cite{Fukushima:2010bq}.
The spin degeneracy will give us a guiding principle for analysis of multi-hadron systems with a single heavy quark.

{\it Acknowledgements.} 
We thank to Prof.~M.~Oka and Prof.~D.~Jido for fruitful discussions.
This work is supported in part by Grant-in-Aid for Scientific Research on 
Priority Areas ``Elucidation of New Hadrons with a Variety of Flavors 
(E01: 21105006)" (S.~Y. and A.~H.) and by Grant-in-Aid for Scientific Research from MEXT
(Grants No. 24105702, No. 24740152 (T.~H.) and No. 22740174 (K.~S.)) and from JSPS (Grants No. 24-3518 (Y.~Y.) and No. 15-5858 (S.~O.)).


\begin{thebibliography}{10}

\bibitem{Brambilla:2010cs}
  N.~Brambilla {\it et al.},
  Eur.\ Phys.\ J.\  C {\bf 71} (2011) 1534.

\bibitem{Cho:2010db}
  S.~Cho {\it et al.} (ExHIC Collaboration),
  Phys.\ Rev.\ Lett.\  {\bf 106} (2011) 212001; Phys.\ Rev.\  C {\bf 84} (2011) 064910.

\bibitem{Isgur:1989vq} 
  N.~Isgur and M.~B.~Wise, 
  Phys.\ Lett.\ B {\bf 232} (1989) 113; ibid. {\bf 237} (1990) 527.


\bibitem{Isgur:1991wq} 
  N.~Isgur and  M.~B.~Wise,
  Phys.\ Rev.\ Lett.\  {\bf 66} (1991) 1130.

\bibitem{Rosner:1985dx} 
  J.~L.~Rosner,
  Comments Nucl.\ Part.\ Phys.\  {\bf 16} (1986) 109.

\bibitem{Eichten:1993ub} 
  E.~J.~Eichten, C.~T.~Hill, and C.~Quigg,
  Phys.\ Rev.\ Lett.\  {\bf 71} (1993) 4116.

\bibitem{Manohar:2000dt}
  A.~V.~Manohar and M.~B.~Wise,
  Camb.\ Monogr.\ Part.\ Phys.\ Nucl.\ Phys.\ Cosmol.\  {\bf 10} (2000) 1.

\bibitem{Flynn:1992fm} 
  J.~M.~Flynn and  N.~Isgur,
  J.\ Phys.\ G {\bf 18} (1992) 1627.


\bibitem{Bardeen:2003kt} 
  W.~A.~Bardeen, E.~J.~Eichten, and C.~T.~Hill,
  Phys.\ Rev.\ D {\bf 68} (2003) 054024.

\bibitem{Matsuki:2007zza} 
  T.~Matsuki, T.~Morii, and K.~Sudoh,
  Prog.\ Theor.\ Phys.\  {\bf 117} (2007) 1077.
  
\bibitem{Roberts:2007ni} 
  W.~Roberts and M.~Pervin,
  Int.\ J.\ Mod.\ Phys.\ A {\bf 23} (2008) 2817.


\bibitem{Yasui:2009bz}
  S.~Yasui and K.~Sudoh,
  Phys.\ Rev.\  D {\bf 80} (2009) 034008.

\bibitem{Yamaguchi:2011xb}
  Y.~Yamaguchi, S.~Ohkoda, S.~Yasui, and A.~Hosaka,
  Phys.\ Rev.\  D {\bf 84} (2011) 014032; ibid. {\bf 85} (2012) 054003.


\bibitem{Yasui:2012rw} 
  S.~Yasui and K.~Sudoh,
  Phys.\ Rev.\ C {\bf 87} (2013) 015202.

\bibitem{Yasui:2013xr} 
  S.~Yasui and K.~Sudoh,
  Phys.\ Rev.\ C {\bf 88} (2013) 015201.

\bibitem{Yamaguchi:2013ty} 
  Y.~Yamaguchi, S.~Ohkoda, S.~Yasui, and A.~Hosaka,
  Phys.\ Rev.\ D {\bf 87} (2013) 074019.


\bibitem{Liu:2011xc} 
  Y.~-R.~Liu and  M.~Oka,
  Phys.\ Rev.\ D {\bf 85} (2012) 014015.

\bibitem{Bayar:2012dd} 
  M.~Bayar, C.~W.~Xiao, T.~Hyodo, A.~Dote, M.~Oka, and E.~Oset,
  Phys.\ Rev.\ C {\bf 86} (2012) 044004.

\bibitem{Mizutani:2006vq} 
  T.~Mizutani and A.~Ramos,
  Phys.\ Rev.\ C {\bf 74} (2006) 065201.

\bibitem{Fukushima:2010bq} 
  K.~Fukushima and T.~Hatsuda,
  Rept.\ Prog.\ Phys.\  {\bf 74} (2011) 014001.
  
\end{thebibliography}

\end{document}